\documentclass[11pt]{article} 
\usepackage{graphicx}
\usepackage{subfigure}
\usepackage{hyperref}

\begin{document} 
\title{Vortex Shedding in the Wake \\ of a Dual Step Cylinder} 
\author{Chris Morton and Serhiy Yarusevych \\ 
\\\vspace{6pt} Department of Mechanical and Mechatronics Engineering, 
\\ University of Waterloo, Waterloo, Ontario, Canada} 
\maketitle 

\begin{abstract} 
A dual-step cylinder is comprised of a large diameter cylinder (D) with low aspect ratio (L/D) attached co-axially to the mid-span of a small diameter cylinder (d). The fluid dynamics video presented in this investigation is used to illustrate the effect of aspect ratio on dual step cylinder wake development for Re$_{D}$ = 2100, D/d = 2, and ${0.2\le L/D\le 3}$. In addition, the video provides visualization of such flow phenomena as interaction of spanwise vortices, development of streamwise vortex filaments, and formation of Kelvin-Helmholtz rollers. The experiments were performed in a water flume at the University of Waterloo. A hydrogen bubble flow visualization technique was employed to visualize vortical structures downstream of each cylinder model. High-resolution images of the flow were obtained with a high speed Photron camera and post-processed using Adobe Photoshop CS4. The results show a plethora of vortices developing in the wake of the dual step cylinder. For ${1\le L/D\le 3}$, spanwise vortex shedding occurs from the large and small cylinders at distinct frequencies. However, for smaller L/D, large cylinder vortex shedding ceases and small cylinder vortices interact in the large cylinder wake.
\end{abstract} 
\section{Introduction} 
Flow around cylindrical structures is common in mechanical and civil engineering applications (e.g., underwater cables, civil support structures, and heat exchangers). One of the most commonly utilized geometries in these applications is a uniform circular cylinder - a geometry that has been investigated extensively in the past (e.g., [1-4]). The present investigation is focused on a dual step cylinder geometry, which has been the focus of far fewer experimental investigations [5,6]. The geometry of interest is shown in Fig. 1, with a large cylinder of diameter D and length L attached at the mid-span of a small cylinder of diameter d. At a sufficiently high Reynolds number, vortex shedding is observed in the wake, however, the wake topology is influenced significantly by the geometrical parameters of the model [5,6]. This article describes a video compilation of turbulent wake visualizations for low aspect ratio dual step cylinders. The main goal of the video is to illustrate the influence of the large cylinder aspect ratio on the turbulent wake development by identifying changes in the formation and development of coherent structures in the wake of the large and small cylinders. 

\section{Experimental Methodology} 
Experiments were completed in a water flume at the University of Waterloo. The test section of the flume is 2.4m long, with a cross section of 1.2m x 1.2m. Dual step cylinder models are constructed of stainless steel and aluminum. A solid smaller diameter cylinder (d) is made of stainless steel while larger diameter cylinders (D) are made of aluminum with holes bored through the center, enabling a concentric assembly with the smaller cylinder. The large diameter cylinders are placed at the mid-span of the smaller diameter cylinder, forming the dual step cylinder configuration depicted in Fig. 1. All of the models were sanded with 1600 grit sand paper and finished with aluminum polish. Hydrogen bubbles were produced along a thin (0.09 mm in diameter) stainless steel wire mounted approximately 0.2D upstream of the model along its entire span, similar to the arrangement employed by Kappler et al. [7] for a uniform circular cylinder. The flow was illuminated with a conical laser beam generated by a two-watt continuous wave Argon-Ion laser and two consecutive cylindrical lenses. The size of the hydrogen bubbles and, hence, their buoyancy depend on the wire diameter and the applied voltage [8]. An experimental study was conducted to determine the DC voltage which produces a sheet of hydrogen bubbles with negligible buoyancy effects. For all hydrogen bubble flow visualization experiments, a DC voltage of fifteen volts was used. Images of the flow were obtained with a high-speed Photron camera. The camera acquired images at 100 Hz, more than twenty times the vortex shedding frequency of small cylinder vortices.

\begin{figure}
\begin{center}
\includegraphics[height=2in]{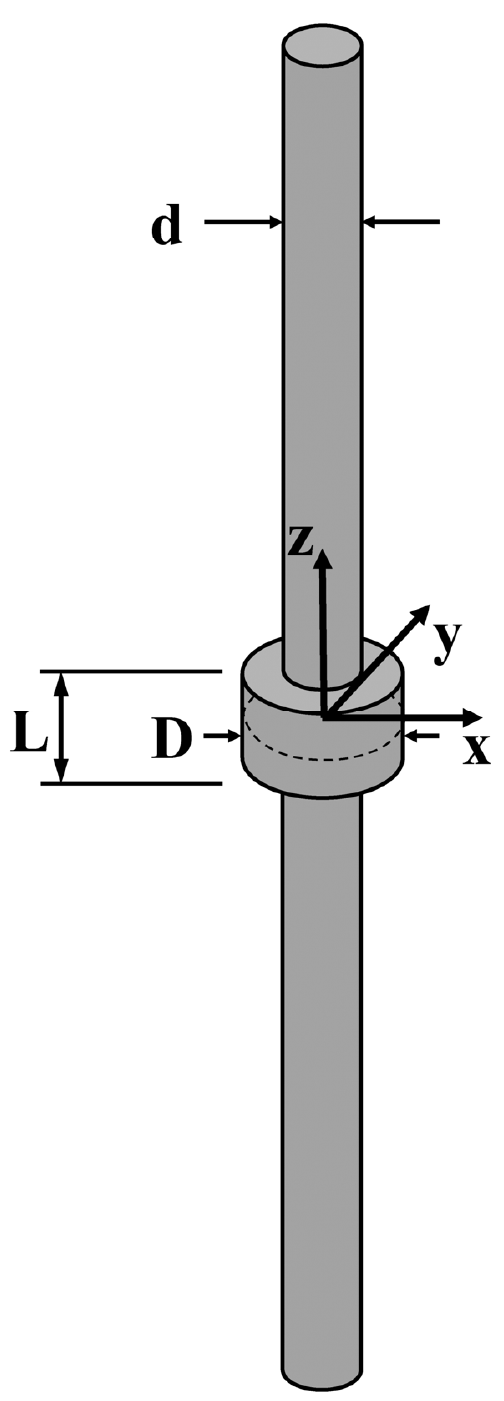}
\caption{Dual step cylinder geometry.}
\end{center}
\end{figure}

\section{Image Processing}
All images were processed with Adobe Photoshop CS4. The raw monochrome images were post processed by applying a gradient contrast adjustment to balance the brightness and a custom gradient map to color the images. In addition, a noise filter was applied to reduce the brightness of sediment particles in the images and a plastic wrap artistic effect was used to enhance the visualization of the vortical structures.

\section{Video Contents}
The video shows the development of wake vortices for dual step cylinder models with L/D = 2, 1, 0.5, and D/d = 2. All the tests were performed at Re$_{D}$ = 2100. Visualization images corresponding to each of the three models investigated are shown in Fig. 2 to aid in the description of the video contents. In the video, specific features of the flow development are highlighted for each L/D, while the effect of L/D is apparent through a comparison of videos. For L/D = 1 (Fig. 2b), vortex shedding is observed in the wake of the large and small cylinders at distinct frequencies. The difference in shedding frequency of the small and large cylinder vortices results in complex vortex interactions (i.e., vortex deformations and vortex splitting). Additionally, at the Reynolds number investigated, an elliptic instability of spanwise vortices leads to the formation of streamwise vortical structures [3]. Footprints of these streamwise vortices appear as mushroom shaped structures in the visualization videos. For L/D = 2 (Fig. 2a), the large cylinder wake topology changes drastically. It is difficult to identify the spanwise vortices shed from the large cylinder, and there is a new phenomenon occurring in the wake: Kelvin-Helmholtz rollers develop downstream of the large cylinder ends. Although they are identifiable in Fig. 2a, the Kelvin-Helmholtz vortices appear and disappear intermittently. These structures are believed to be formed due to the roll-up of the shear layers emanating from the steps. For L/D = 0.5 (Fig. 2c), vortex shedding from the large cylinder ceases. The video shows that the large cylinder separated shear layers collapse within a few diameters downstream of the model. Downstream of this location, some small cylinder vortices form connections across the wake of the large cylinder while others form half-loop vortex connections, as depicted in Fig. 2c.

\begin{figure}
\begin{center}
\subfigure[L/D = 2]{\includegraphics[height=2.5in]{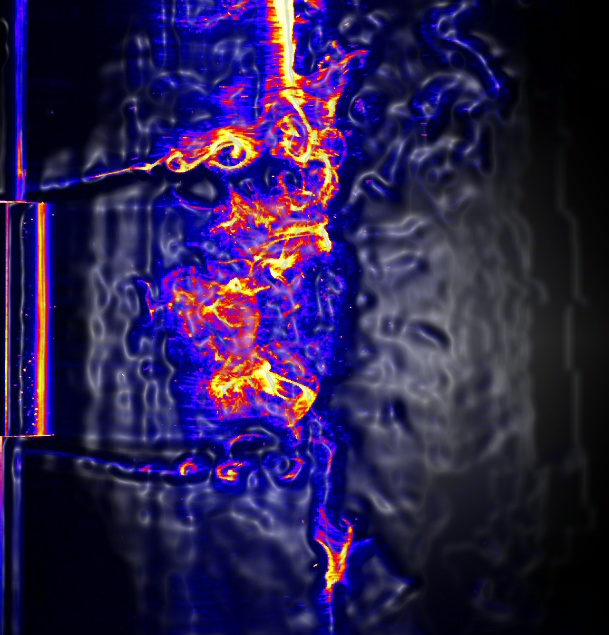}
}
\subfigure[L/D = 1]{\includegraphics[height=2.5in]{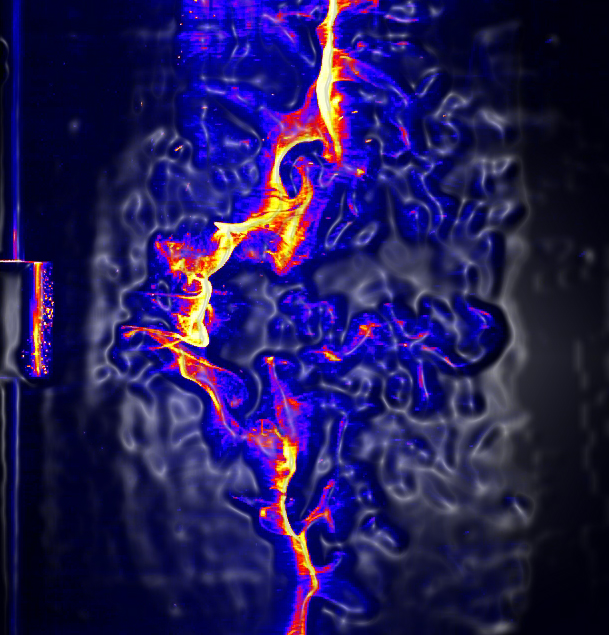}
}
\subfigure[L/D = 0.5]{\includegraphics[height=2.5in]{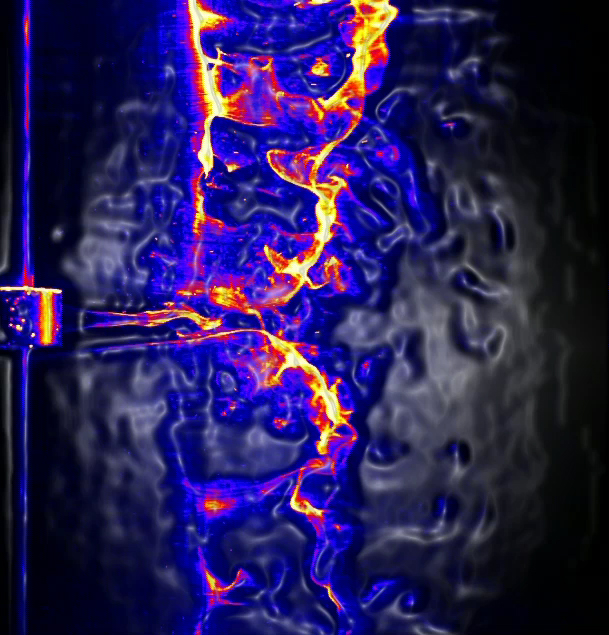}
}
\caption{Hydrogen bubble visualization of vortex shedding in the wake of a dual step cylinder.}
\end{center}
\end{figure}

\section{Acknowledgements}
The authors gratefully acknowledge the Natural Sciences and Engineering Research Council of Canada (NSERC) for funding of this work.

\section{References}
[1] Gerrard, J.H. (1978) “The wakes of cylindrical bluff bodies at low Reynolds number,” Philosophical Transactions of the Royal Society of London, Vol. 288, pp. 351-382.\newline
[2] Roshko, A. (1993) “Perspectives on bluff body aerodynamics”, Journal of Wind Engineering and Industrial Aerodynamics, Vol. 49, pp. 79-100.\newline
[3] Williamson, C.H.K. (1996) “Vortex dynamics in the cylinder wake,” Annual Review of Fluid Mechanics, Vol. 28, pp. 477-539.\newline
[4] Norberg, C. (2003) “Fluctuating lift on a circular cylinder: review and new measurements,” Journal of Fluids and Structures, Vol. 17, pp. 57-96.\newline
[5] Williamson, C.H.K. (1992) “The natural and forced formation of spot-like ‘vortex dislocations’ in the transition of a wake,” Journal of Fluid Mechanics, Vol. 243, pp.  393-441.\newline
[6] Morton, C., and Yarusevych, S. (2011) “An experimental investigation of flow past a dual step cylinder,” Experiments in Fluids, DOI 10.1007/s00348-011-1186-z.\newline
[7] Kappler, M., Rodi, W., Szepessy, S., and Badran, O. (2005) “Experiments on the flow past long circular cylinders in a shear flow,” Experiments in Fluids, Vol. 38, pp. 269-284.\newline
[8] Lian, Q.X., and Su, T.C., (1996) “The Application of Hydrogen Bubble Method in the Investigation of Complex Flows,” Atlas of Flow Visualization II, CRC Press, Chap. 6.\newline
\end{document}